\documentclass[aps,prb,twocolumn,groupedaddress,showpacs,floatfix,altaffilletter]{revtex4-1}
\usepackage[pdftex,plainpages=false,colorlinks=true,linkcolor=blue, citecolor=blue, urlcolor=blue]{hyperref}
\usepackage{amsfonts}
\usepackage{amsmath}
\usepackage{amssymb}
\usepackage{array}
\usepackage{graphicx}% Include figure files
\usepackage{natbib}

\newcolumntype{L}[1]{>{\raggedright\let\newline\\\arraybackslash\hspace{0pt}}m{#1}}
\newcolumntype{C}[1]{>{\centering\let\newline\\\arraybackslash\hspace{0pt}}m{#1}}
\newcolumntype{R}[1]{>{\raggedleft\let\newline\\\arraybackslash\hspace{0pt}}m{#1}}

\begin{document}
\title{Ergodicity of the Hybridization-Expansion Monte Carlo Algorithm for Broken-Symmetry States}

\author{P. S\'emon}
\affiliation{D\'{e}partement de physique and Regroupement qu\'{e}b\'{e}cois sur les mat\'{e}riaux de pointe, Universit\'{e} de Sherbrooke, Sherbrooke, Qu\'{e}bec, Canada J1K 2R1}
\author{G. Sordi}
\affiliation{SEPnet and Hubbard Theory Consortium, Department of Physics, Royal Holloway, University of London, Egham, Surrey, UK, TW20 0EX}
\author{A.-M. S. Tremblay}
\affiliation{D\'{e}partement de physique and Regroupement qu\'{e}b\'{e}cois sur les mat\'{e}riaux de pointe, Universit\'{e} de Sherbrooke, Sherbrooke, Qu\'{e}bec, Canada J1K 2R1}
\affiliation{Canadian Institute for Advanced Research, Toronto, Ontario, Canada, M5G 1Z8}

\begin{abstract}
With the success of dynamical mean field theories, solvers for quantum-impurity problems have become an important tool for the numerical study of strongly correlated systems. Continuous-time Quantum Monte Carlo sampling of the expansion in powers of the hybridization between the ``impurity'' and the bath provides a powerful solver when interactions are strong. Here we show that the usual updates that add or remove a pair of creation-annihilation operators are rigorously not ergodic for several classes of broken-symmetries that involve spatial components. We show that updates with larger numbers of simultaneous updates of pairs of creation-annihilation operators remedy this problem. As an example, we apply the four operator updates that are necessary for ergodicity to the case of d-wave superconductivity in plaquette dynamical mean-field theory for the one-band Hubbard model. While the results are qualitatively similar to those previously published, they are quantitatively better that previous ones, being closer to those obtained by other approaches. 

\end{abstract}


\maketitle

\section{Introduction}
Understanding and predicting the different phases of matter is one of the main goals of condensed matter physics. Some phases break symmetries of the underlying Hamiltonian. This can happen in an infinite system only. Mean field theories are an important tool for the study of broken symmetries since the infinite system limit is naturally taken into account. While ordinary mean field theories are sufficient for weakly correlated systems, they fail for strongly correlated systems such as doped Mott insulators~\cite{ift}, high temperature superconductors,~\cite{lee,Scalapino_RMP:2012,TremblayJulichPavarini:2013} layered organic superconductors~\cite{powell_strong_2006,PowellMcKenzieReview:2011} and the like. Here dynamical mean field theories~\cite{rmp,maier,tremblayR} are necessary for an adequate treatment. They self-consistently map the infinite lattice model on a quantum-impurity model consisting of a finite interacting system immersed in a non-interacting electronic bath.

A breakthrough in the solution of quantum-impurity problems has occurred with the advent of Continuous-Time Quantum Monte-Carlo algorithms (CTQMC).~\cite{millisRMP} These algorithms come in various guises: For example, the Rubtsov algorithm,~\cite{Rubtsov:2005} auxiliary-field algorithm~\cite{gull2008continuous} and the hybridization expansion algorithm~\cite{Werner:2006General,Werner:2006,hauleCTQMC}. Here we focus on the latter algorithm (CT-HYB) that is especially suited at strong coupling~\cite{gullCOMP} and for ab-initio codes that are combined with dynamical mean-field theory.~\cite{kotliarRMP} 

We show that for several classes of broken symmetries that involve spatial components, CT-HYB is not ergodic as a matter of principle if one follows the standard update procedure of adding or removing a single pair of creation-annihilation operators. This deficiency can be cured by updates that add more pairs of creation-annihilation operators. As an important example, we consider the case of d-wave superconductivity on the square lattice that breaks not only $U(1)$ symmetry but also rotation by $\pi/2$. The solution of the quantum-impurity problem consisting of the Hubbard model on a plaquette immersed in a bath is made self-consistent with the lattice problem through Cellular Dynamical-Mean-Field theory~\cite{gabiCDMFT}. The resulting phase diagram is qualitatively similar with the previously published one~\cite{Sordi:2012SC} but quantitatively more reliable since in the zero-temperature limit the range of doping where superconductivity appears agrees with results obtained with the exact-diagonalization impurity solver~\cite{kancharla}.

In Sec.~\ref{Sec:ImpurityModel} we introduce an effective quantum-impurity model for a correlated problem on an infinite lattice, along with the self-consistency condition for Cellular-Dynamical Mean-Field theory (CDMFT). All of our formal results on Monte Carlo updates apply to the hybridization expansion, whatever the self-consistency condition between impurity and lattice. We then recall in Sec.~\ref{Sec:HybridizationExpansion} the general formalism for the CT-QMC hybridization solver. The question of ergodicity is discussed in Sec.~\ref{Sec:ErgodicUpdates}. After demonstrating in the first subsection why standard updates with pairs of creation-annihilation operators are not ergodic using the example of d-wave superconductivity, we show how updates with two pairs of creation-annihilation operators solve the problem for this case. The phase diagram is discussed in the following subsection while the case of a general broken spatial symmetry is addressed in the last subsection. We conclude in Sec.~\ref{Sec:Discussion}

\section{Effective impurity model}\label{Sec:ImpurityModel}
%Within CDMFT, let us consider the effective problem of a cluster, also called ``impurity'', immersed in a bath. With $H_\text{loc}$ the Hamiltonian of the interacting system (\textcolor{red}{connection between interacting system and impurity missing}), 
The effective quantum-impurity problems we are interested in consists of an interacting system, described by $H_{\text{loc}}(d_i^\dagger, d_i)$, immersed in a non-interacting bath. The Hamiltonian for the impurity plus bath takes the form
\begin{equation}
\label{equ:imp}
\begin{split}
H_{\text{imp}} &= H_{\text{loc}}(d_i^\dagger, d_i) + \sum_{i\mu}(V_{\mu i} a_{\mu}^\dagger d_{i} + V_{\mu i}^* d_{i}^\dagger a_\mu) \\
&\quad\quad + \sum_{\mu} \epsilon_{\mu} a_{\mu}^\dagger a_{\mu},
\end{split}
\end{equation} 
with $\epsilon_{\mu}$ the bath dispersion and $V_{\mu i}$ the amplitude for a particle to hop from the system orbital $i$ to the bath orbital $\mu$. We include spin and position in the definition of impurity orbitals. The self-energy $\Sigma$ of this impurity problem is finite for the interacting system only, so that when the bath is integrated out, Dyson's equation takes the form
\begin{equation}
G_{\text{loc}}^{-1} = G_{0,\text{loc}}^{-1} - \Delta - \Sigma,
\end{equation}
where $G_{\text{loc}}^{-1}$ and $G_{0,\text{loc}}^{-1}$ are the interacting and non-interacting cluster Green's functions respectively. The bath degrees of freedom are encapsulated in the hybridization function  
\begin{equation}
\label{equ:hyb}
\Delta_{i j}(i\omega_n) =  \sum_{\mu} \frac{V^*_{\mu i} V_{\mu j}}{i\omega_n - \epsilon_{\mu}},
\end{equation}
which plays the role of the dynamical mean field.%\newline

For the self-consistent mapping between the lattice and impurity, CDMFT~\cite{gabiCDMFT} starts with a periodic partitioning of the lattice system into disconnected clusters. Taking for $H_{\text{loc}}$ the restriction of the lattice Hamiltonian to one of these clusters and representing the rest of the lattice by a non-interacting bath, the hybridization function is self-consistently obtained from a restriction of the lattice Dyson equation
\begin{equation}
G_{\text{loc}}[\Delta]=(G_{0,\text{latt}}^{-1} - \Sigma_{\text{latt}}'[\Delta])^{-1}|_{\text{loc}}
\end{equation}
to the cluster, with $G_{0,\text{latt}}$ the non-interacting lattice Green's function. The approximate lattice self-energy $\Sigma_{\text{latt}}'$  
%is taken equal to that found from the solution of the impurity model.
equals the impurity-model self-energy on everyone of the clusters. 
%on every plaquette of the lattice. \newline

This self-consistent mapping on an impurity problem conserves the symmetries of the lattice system compatible with the partitioning. In the normal phase, the dynamical mean field is constrained to satisfy these symmetries, while in a broken symmetry phase it is allowed to break some of them. The symmetry is thus broken in the dynamical mean fields and not on the cluster. This applies to the dynamical cluster approximation $\text{DCA}$ as well.~\cite{Hettler:1998} %\newline 

In order to satisfy the self-consistency condition, CDMFT and DCA require an infinite number of bath orbitals. Only CTQMC impurity solvers give (statistically) exact solutions in this limit. The CT-HYB impurity solver of interest here is reviewed in the next section. 

\section{Hybridization expansion for Continuous-Time Quantum Monte Carlo}\label{Sec:HybridizationExpansion} 

This summary of the CT-HYB algorithm~\cite{Werner:2006General,Werner:2006,hauleCTQMC,millisRMP} focuses on the aspects relevant for the rest of the discussion on ergodicity. First, the impurity Hamiltonian is rearranged as  
\begin{equation}
H_\text{imp}=H_\text{loc} + H_{\text{hyb}} + H_{\text{hyb}}^\dagger + H_\text{bath},
\end{equation}
where $H_\text{bath} =\sum_\mu \epsilon_{\mu} a_{\mu}^\dagger a_{\mu}$ and $H_{\text{hyb}} = \sum_{i\mu} V_{\mu i} a_{\mu}^\dagger d_{i}$. Writing the impurity partition function $Z=\text{Tr}e^{-\beta H_{\text{imp}}}$ in the interaction representation and expanding in powers of the hybridization term yields 
\begin{equation}
\begin{split}
Z &= \text{Tr}\text{T}_\tau e^{-\beta H_0} e^{ -\int_0^\beta d\tau  (H_{\text{hyb}}(\tau) + H_{\text{hyb}}^\dagger(\tau))}\\
&= \sum_{k\ge 0}\frac{1}{(2k)!} \int_0^\beta d\tau_1\cdots\ d\tau_{2k}\text{Tr}\text{T}_\tau e^{-\beta H_0} \bigl (H_\text{hyb}(\tau_1) \\
&\quad+ H_\text{hyb}^\dagger(\tau_1)\bigr) \cdots \bigl ( H_\text{hyb}(\tau_{2k}) + H_\text{hyb}^\dagger(\tau_{2k})\bigr)\\
&= \sum_{k\ge 0}  \frac{1}{k!^2} \int_0^\beta d\tau_1 \cdots d\tau_k \int_0^\beta d\tau_1'\cdots d\tau_k' \text{Tr}\text{T}_\tau e^{-\beta H_0}\\
&\quad \times H_\text{hyb}(\tau_1)H_\text{hyb}^\dagger(\tau_1') \cdots H_\text{hyb}(\tau_{k})H_\text{hyb}^\dagger(\tau_{k}').
\end{split}
\end{equation}
As $H_\text{loc}$ conserves the particle number, odd expansion orders vanish and there are $(2k)!/k!^2$ finite terms when multiplying out the second line. Defining $\hat{V}_i = \sum_\mu V_{\mu i}^* a_\mu$ and replacing the hybridization terms, the cluster and bath degrees of freedom are separated 
\begin{equation}
\label{equ:partitionexpansion}
\begin{split}
Z &= \sum_{k\ge 0} \sum_{i_1\cdots i_k} \sum_{i_1' \cdots i_k'} \frac{1}{k!^2} \int_0^\beta d\tau_1 \cdots d\tau_k \int_0^\beta d\tau_1'\cdots d\tau_k' \\
&  \quad \times \text{Tr}\text{T}_\tau e^{-\beta H_0} \hat{V}_{i_1}^\dagger(\tau_1) d(\tau_1) \cdots d^\dagger(\tau_k') \hat{V}_{i_k}(\tau_k') \\
&= \sum_{k\ge 0} \sum_{i_1\cdots i_k} \sum_{i_1' \cdots i_k'} \frac{1}{k!^2} \int_0^\beta d\tau_1 \cdots d\tau_k \int_0^\beta d\tau_1'\cdots d\tau_k' \\
& \quad  \times \text{Tr}\text{T}_\tau e^{-\beta H_{\text{loc}}} d_{i_1}(\tau_1) d_{i_1'}^\dagger(\tau_1')\cdots  d_{i_k}(\tau_k) d_{i_k'}^\dagger(\tau_k') \\
&\quad \times Z_{\text{bath}} \langle \hat{V}^\dagger_{i_1}(\tau_1)\hat{V}_{i_1'}(\tau_1')\cdots  \hat{V}^\dagger_{i_k}(\tau_k) \hat{V}_{i_k'}(\tau_k') \rangle,
\end{split}
\end{equation}
where $\langle O \rangle := Z^{-1}_\text{bath}\text{Tr}[\text{T}_\tau e^{-\beta H_\text{bath}} O]$ and $Z_\text{bath}$ is the bath partition function. %\newline

\begin{figure}[t]
\includegraphics[width=0.9\linewidth, clip=]{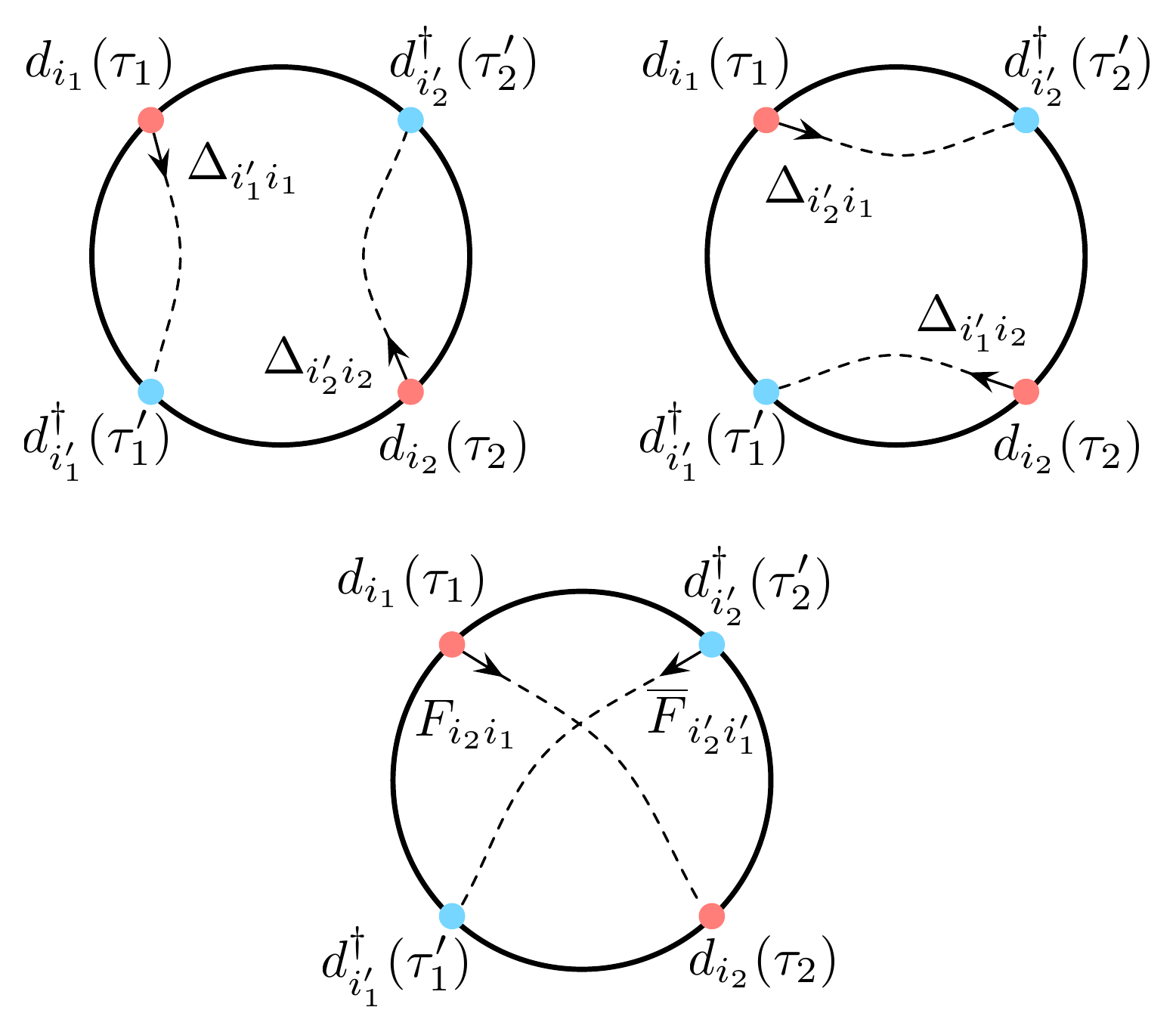} 
\caption{Diagrams contributing to the weight of a second order configuration, c.f. Eq.~(\ref{equ:Wick}). The bold black circle represents the trace with the impurity operators, connected in all different ways by the hybridization function.}
% (only two of the three possible contractions are shown here).}
\label{fig11}
\end{figure}

The bath is quadratic, and with Wick's theorem the average over the bath is expressed as a sum over all contractions, e.g. at second order
\begin{equation}
\label{equ:Wick}
\begin{split}
&\langle \hat{V}_{i_1}^\dagger(\tau_1)\hat{V}_{i'_1}(\tau_1')\hat{V}_{i_2}^\dagger(\tau_2) \hat{V}_{i_2'}(\tau_2')\rangle = \\
&\quad \langle\hat{V}_{i_1}^\dagger(\tau_1)\hat{V}_{i_1'}(\tau_1')\rangle \langle\hat{V}_{i_2}^\dagger(\tau_2)\hat{V}_{i_2'}(\tau_2')\rangle -\langle\hat{V}_{i_1}^\dagger(\tau_1)\hat{V}_{i_2'}(\tau_2')\rangle \\
&\quad \times\langle\hat{V}_{i_2}^\dagger(\tau_2)\hat{V}_{i_1'}(\tau_1')\rangle -\langle\hat{V}_{i_1}^\dagger(\tau_1)\hat{V}_{i_2}^\dagger(\tau_2)\rangle \langle\hat{V}_{i_1'}(\tau_1')\hat{V}_{i_2'}(\tau_2')\rangle , 
\end{split}
\end{equation}
where $\langle\hat{V}_{i}^\dagger(\tau)\hat{V}_{i'}(\tau')\rangle$ evaluates to the hybridization function $\Delta_{i'i}(\tau' - \tau)$ in Eq.~\eqref{equ:hyb}. The anomalous hybridization functions $F_{i_2i_1}(\tau_2 - \tau_1) := \langle\hat{V}^\dagger_{i_1}(\tau_1)\hat{V}^\dagger_{i_2}(\tau_2)\rangle$ and $\overline{F}_{i_2i_1}(\tau_2 - \tau_1) := \langle\hat{V}_{i_2}(\tau_1) \hat{V}_{i_1}(\tau_2)\rangle$ vanish for a particle number conserving bath as in Eq.~(\ref{equ:imp}). A contraction may be represented as shown in Fig.~\ref{fig11}, and the sum over all finite contractions can in most cases be cast into a determinant.%\newline

In Quantum Monte Carlo one interprets the terms of the series (\ref{equ:partitionexpansion}), supposed positive here for simplicity, as weights $w$ for a probability distribution $w/Z$ over the configuration space $\mathcal{C}:=\lbrace (\tau_1 i_1\,\tau_1'i_1' \dots \tau_ki_k\,\tau_k'i_k') | k \ge 0 \rbrace$. Observables, such as the local Green's function, can be expressed as random variables over $\mathcal{C}$. To obtain estimates, the probability distribution is sampled by a Markov process $c_1 \rightarrow c_2 \rightarrow \dots$ in $\mathcal{C}$, characterized by the transition probability $P(c_{i + 1}|c_i)$ of going from configuration $c_i$ to configuration $c_{i + 1}$. The Markov process converges to $w/Z$ if the transition probability satisfies detailed balance $P(c_{i+1}|c_i)w(c_i) = P(c_i|c_{i+1})w(c_{i+1})$ and ergodicity.%\newline 

The Metropolis-Hasting algorithm gives a possible choice for the transition probability. To start with a trial configuration $c$ is chosen according to a trial probability $q(c|c_i)$, and we set $c_{i+1}:=c$ with probability   
\begin{equation}
p=\text{min}\biggl (\frac{q(c_i|c)w(c)}{q(c|c_i)w(c_i)}, 1 \biggr)
\end{equation}
and $c_{i + 1} := c_i$ otherwise. This transition probability $p\cdot q$ satisfies detailed balance. \newline

\section{Ergodic updates in the presence of broken symmetry}\label{Sec:ErgodicUpdates} 

\subsection{Standard updates}\label{SubSec:StandardUpdates}
For an ergodic Metropolis-Hasting sampling, the transition probability should allow to explore all the configuration space. With respect to the trial probability, this sets two conditions.%\newline

First, the proposed updates should allow to go from any configuration to any configuration. A natural choice here is the insertion or the removal of two impurity operators $d_{i}(\tau) d_{i'}^\dagger(\tau')$. 
Second, the weights of the configurations along the proposed path have to be finite. For some configurations, the trace may vanish due to symmetry constraints. If this happens along all paths between two configurations, the two operator updates are not ergodic. This is illustrated in the next section.
\subsection{Updates for ergodicity in the presence of superconductivity}\label{SubSec:UpdatesForSC}
%(devrais-je mentionner qu'on utilise Nambu ? Car c'est plus facile de fixer l'ordre des spin au lieu de les laisser libre).
Consider a CDMFT study of d-wave superconductivity in the 2D Hubbard model with a 2x2 cluster. As the cluster Hamiltonian conserves, beside charge and spin $\sigma$, the cluster momentum $\mathbf{K}\in\lbrace (0,0),(\pi,0),(0,\pi),(\pi,\pi)\rbrace$, it is numerically advantageous to label the one particle basis by $\mathbf{K}$.~\cite{hauleCTQMC}%\newline 

In the normal phase only  the diagonal hybridization entries $\Delta_{\sigma \mathbf{K} , \sigma \mathbf{K}}$ are finite. In the superconducting phase charge conservation is broken, and the anomalous entries $F_{\uparrow \mathbf{K}, \downarrow -\mathbf{{K}}}$ as well as their conjugates  $\overline{F}_{\uparrow \mathbf{K}, \downarrow -\mathbf{{K}}}$ may be finite. The d-wave order parameter changes sign under rotation by $\pi/2$ and hence $F_{\uparrow (0,\pi),\downarrow (0,\pi)}=-F_{\uparrow (\pi,0), \downarrow (\pi,0)}$ while $F_{\uparrow (0,0), \downarrow(0,0)}$ and $F_{\uparrow (\pi,\pi), \downarrow (\pi,\pi)}$ vanish.%\newline

Only insertions or removals of $d_{\sigma \mathbf{K}}^\dagger d_{\sigma \mathbf{K}}$ operators lead to a finite trace since $\mathbf{K}$ is conserved. Hence, starting from expansion order zero, the two operator updates only reach configurations where for each $\sigma,\mathbf{K}$ there is the same number of $d_{\sigma \mathbf{K}}^\dagger$ and $d_{\sigma \mathbf{K}}$. The finite second order configuration 
\begin{equation}
\label{equ:SCConfig}
\begin{split}
&\text{Tr}[d_{\uparrow (0,\pi)}d_{\downarrow (0,\pi)}d^\dagger_{\downarrow (\pi,0)}d^\dagger_{\uparrow (\pi,0)}]\\
&\quad \quad \quad \times F_{\uparrow (0,\pi),\downarrow (0,\pi)} \overline{F}_{\uparrow (\pi,0), \downarrow (\pi,0)}
\end{split}
\end{equation}
in the superconducting phase does not meet this condition, and the two operator updates are not ergodic. Insertion or removal of these four operators or their conjugates at once is thus a necessary condition for ergodicity. %\newline
 
To show that these four operator updates restore ergodicity in principle, it is sufficient to connect an arbitrary finite configuration to expansion order zero, as this allows to go from any configuration to any configuration by detailed balance. Consider any finite configuration. It can be decomposed into groups of two or four operators which transform as the identity. Groups of two operators come from finite contractions with normal hybridization functions $\Delta_{\sigma \mathbf{K},\sigma \mathbf{K}}$. In addition, by charge conservation on the impurity, all possible anomalous contractions can be grouped in pairs of the form 
%be groups of four operators if there is a certain number of superconducting hybridization functions $F_{\uparrow \mathbf{K}, \downarrow -\mathbf{{K}}}$, by charge conservation on the impurity, there must also be the same number of $\overline{F}_{\uparrow \mathbf{K}', \downarrow -\mathbf{{K}'}}$. We regroup them as 
$F_{\uparrow \mathbf{K}, \downarrow -\mathbf{{K}}}\overline{F}_{\uparrow \mathbf{K}', \downarrow -\mathbf{{K}'}}$, where $\mathbf{{K}'}$ and $\mathbf{{K}}$ can be different. The corresponding group of four operators transforms as the identity, and the four operator updates allows us to remove them. If $\mathbf{K}=\mathbf{K}'$, they may also be removed by two times a two operator update. Hence every configuration can be reached from zero expansion order.%\newline

In the following section, we illustrate how the four operator updates reconcile results obtained with different methods.  
\begin{figure}[!ht]
\includegraphics[width=0.9\linewidth, clip=]{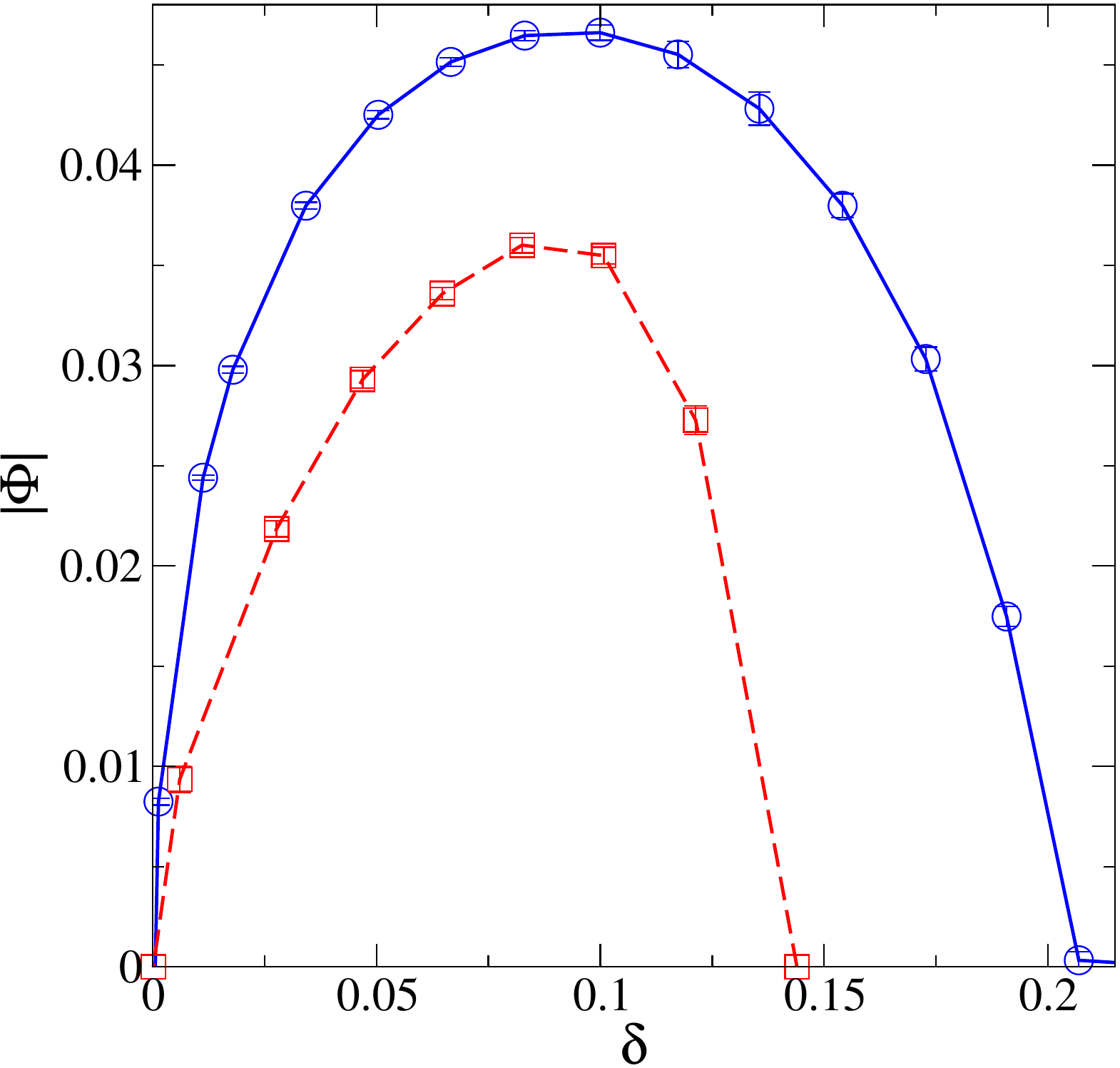}
\caption{$d$-wave superconducting order parameter $\Phi$ as a function of doping $\delta$, for the low temperature $T=1/100$, with and without four operator updates (circles and squares respectively). The value of the interaction $U=9.0$ is larger than $U_{\rm MIT}$. }
\label{fig1}
\end{figure}
\subsection{Numerical results for the superconducting state}
Consider the Hubbard model on a square lattice with on-site interaction $U$ and nearest-neighbor hopping $t$. We follow the notation of Ref.~\onlinecite{Sordi:2012SC} and use CDMFT on a $2\times 2$ plaquette. 

We begin with $U=9.0$, which is above the Mott transition endpoint at half filling $U_{\rm MIT}\approx 5.95$~\cite{sht2,phk}. Figure~\ref{fig1} shows the $d$-wave superconducting order parameter $\Phi$ at the low temperature $T/t =1/100$ as a function of doping, with and without four operator updates (circles and squares, respectively). 
In both cases, $\Phi=0$ in the Mott insulator at zero doping, then it increases upon hole doping, reaches a maximum around $\delta\approx0.09$, and finally it decreases with further doping. 
Notice that the position of the maximum of $\Phi$ remains approximately the same, and it occurs for a doping near the underlying normal state transition between a pseudogap and a correlated metal.~\cite{Sordi:2012SC,sht}

The effect brought about by the four operator updates is twofold: the overall strength of $\Phi$ is larger and $\Phi$ extends over a larger range of dopings when the four operator updates are considered. 
%These two effects suggest that the additional superconducting configurations explored by the four operator updates enhance superconductivity. 
%Note that with the four operator updates the magnitude of $\Phi$ is consistent with the results found with exact diagonalization at $T=0$~\cite{kancharla}
The range of dopings where superconductivity occurs is now consistent with the results found at $T=0$ in Ref.~\onlinecite{kancharla}.

\begin{figure}[t]
\includegraphics[width=0.95\linewidth, clip=]{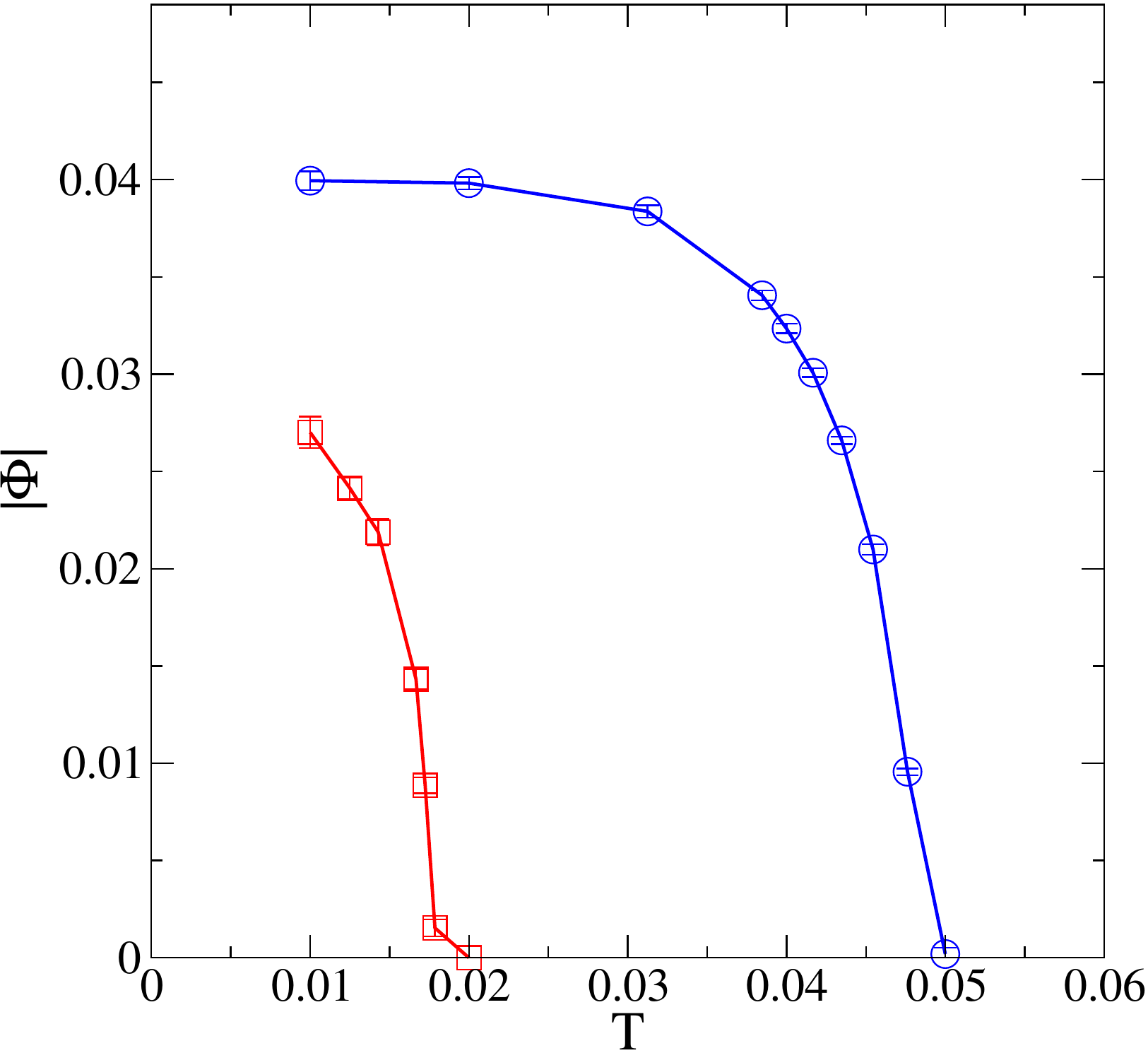} 
\caption{$d$-wave superconducting order parameter $\Phi$ as a function of temperature $T$ for $U=9.0$ and $\delta=0.04$, with and without four operator updates (circles and squares respectively)}
\label{fig2}
\end{figure}

Figure~\ref{fig2} shows the superconducting order parameter $\Phi$ at $\delta=0.04$ as a function of temperature $T$, with and without four operator updates (circles and squares, respectively). 
In both cases, $\Phi$ decreases with increasing $T$ and disappears at the CDMFT transition temperature $T_c^d$. 
We determine $T_c^d$ as the mean of the two temperatures where $\Phi$ changes from finite to zero within error bars. 

Physically, $T_c^d$ is the temperature below which Cooper pairs form within the $2\times2$ plaquette. In Ref.~\onlinecite{Sordi:2012SC} we pointed out that $T_c^d$ is distinct from the pseudogap temperature $T^*$ and can be associated to local pair formation observed in tunnelling spectroscopy~\cite{Gomes:2007,Gomes:2008}. 

\begin{figure}[t]
\includegraphics[width=0.9\linewidth]{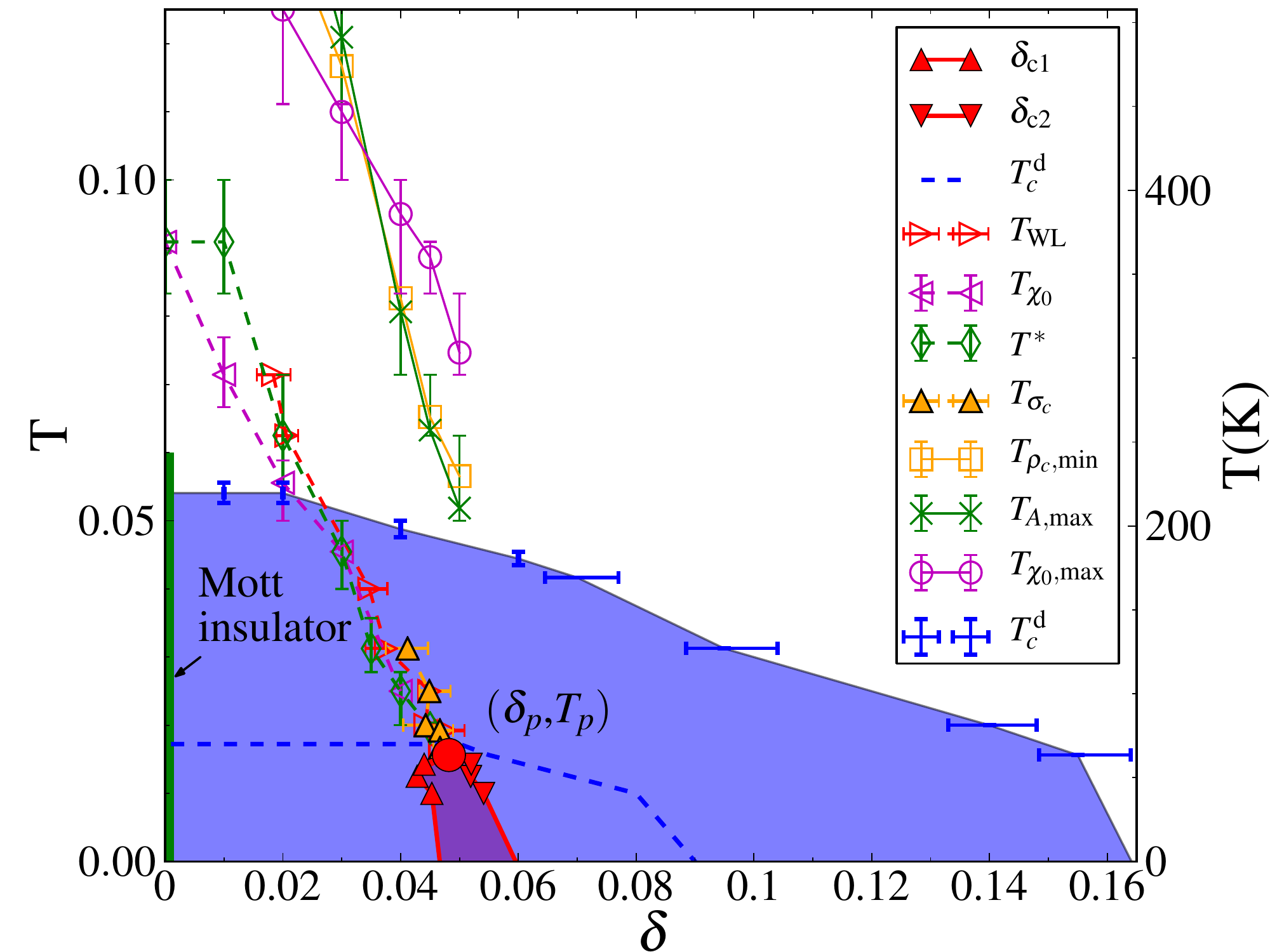} 
\caption{(Color online) Revised temperature versus doping phase diagram of the two dimensional Hubbard model within plaquette CDMFT for $U=6.2$. The only modification compared with Refs.~\onlinecite{Sordi:2012SC, Sordi:2013c} is for the superconducting region delineated by $T_c^d$ (blue/light grey area). With two-operator updates, superconductivity occurs below the dotted blue (light grey) line. With the four-operator updates, superconductivity extends to the end of the blue (light-grey) area. For completeness, we describe the rest of the phase diagram. The first-order transition (red/dark grey area) terminating at the critical endpoint $(\delta_p, T_p)$ (circle) separates a correlated metal from a pseudogap metal.
$T_{\sigma_c}(\delta)$ is the temperature where $\sigma_c(\mu)$ has an inflection point.
It follows $T^*$ and $T_{\rm WL}$, i.e. the dynamic and thermodynamic supercritical crossovers determined by the inflection in the local density of states $A(\omega=0,T)$ and in the charge compressibility $\kappa(\mu)$ respectively~\cite{ssht}. 
The pseudogap scale can be identified also as inflection points in the local spin susceptibility $\chi_0(T)$, $T_{\chi_0}$. 
$T_{\rho_c, \rm min}$ is the temperature where $\rho_c(T)$ has a minimum. It scales with the temperature where $A(\omega=0,T)$ [$\chi_0(T)$] peaks, $T_{A, \rm max}$ [$T_{\chi_0, \rm max}$].
}
\label{fig3}
\end{figure}

Finally, it is important to evaluate the role of the four operator updates on the scenario for the interplay between superconductivity and Mott physics that we have put forward in Refs.~\onlinecite{Sordi:2012SC, Sordi:2013c}. Fig.~\ref{fig3} shows the temperature versus doping phase diagram considered in those references. The value of the interaction is $U=6.2$ and both superconducting and normal state are shown. 

First, let us focus on $T_c^d$, indicated by full and dashed blue line (with and without four operator updates, respectively). The effects brought about by the four operator updates are solely quantitative: the superconducting phase delimited by $T_c^d$ extends over a large range of doping and temperature. The main qualitative features of $T_c^d$ remain however
unchanged: (i) at zero doping, $T_c^d$ is zero, (ii) at all numerically accessible small dopings $T_c^d$ has a finite value, which does not show large variations when a pseudogap appears in the underlying normal state, and (iii) with further doping beyond the pseudogap, $T_c^d$ decreases and eventually vanishes at large doping.

Second, the interplay between superconductivity and Mott physics discussed in earlier papers~\cite{Sordi:2012SC, Sordi:2013c} is still valid. The first-order transition at finite doping separating a pseudogap from a correlated metal is continuously connected to the first-order Mott transition at half-filling.~\cite{sht,sht2} The crossovers lines emerging out of the finite-doping first-order transition signal the appearance of a Mott-driven pseudogap at along a line, $T^*$, at finite temperature~\cite{ssht}. The crossovers intersect the superconducting state delimited by $T_c^d$, implying that pseudogap and superconductivity are distinct phenomena. Superconductivity can emerge either from a pseudogap phase or from a correlated metal, a result confirmed by large cluster studies.~\cite{GullSupercond:2013,GullEnergetics:2012,chen_evolution_2013,sakaiSC} A discussion of the general features of these theoretical results in the context of experiments appears in Ref.~\onlinecite{Alloul:2013}.

Note that since $T_c^d$ is largest for values of $U$ close to $U_{MIT}$, it is comforting that the four operator updates take $T_c^d$ well above $100K$, as shown in Fig.~\ref{fig3}. Indeed, the cuprates are described by a larger $U$ than the one studied here, so calculations will lead to a smaller optimal $T_c^d$. This $T_c^d$ should nevertheless still be above the maximal $T_c$ since it is a mean-field result. Long-wavelength fluctuations and other non mean-field effects can only make the true $T_c$ smaller than $T_c^d$.

\subsection{Updates for ergodicity in the presence of general broken symmetries}\label{SubSec:UpdatesGeneral}

The lack of ergodicity of two-operator updates occurs more generally with broken symmetries. Before we discuss this, let us return to the case of superconductivity.
In the normal phase, configurations which are problematic in the superconducting phase have vanishing weight because the corresponding hybridization functions vanish. 
%The superconducting phase however adds contractions and these configurations become finite. 
The ergodicity of the two operator updates thus depends on the structure of the hybridization function.%\newline 

To render this dependence more explicit, we begin by following the lines of Sec. (\ref{SubSec:UpdatesForSC}), but considering an arbitrary abelian symmetry group $G$ instead of the translation symmetry that gave us conservation of $\mathbf{K}$. 
%Non-abelian groups do not allow products of creation-annihilation operators in the corresponding irreducible representations to establish a unique connection between subspaces.  
%The superconducting order parameter is translationally invariant and only the $F_{\uparrow \mathbf{K}_1, \downarrow \mathbf{K}_2}$ with $\mathbf{K}_1 + \mathbf{K}_2 = 0$ may be finite. In other words, the broken symmetry transforms spatially as an irreducible representation of the translation symmetry, and we proceed likewise for other spatial symmetries. 
Replacing the momenta $\mathbf{K}$ by the characters $\chi$ of $G$, all $F_{\uparrow \chi_1,\downarrow \chi_2}$ with $\chi_1\chi_2=\chi_0$ and their conjugates are allowed to be finite.\footnote{We assume that a character appears at most once in the one particle basis of irreducible representations.} While the configuration  
\begin{equation}
\label{equ:SCConfig2}
\text{Tr}[d_{\uparrow \chi_1}d_{\downarrow \chi_2}d^\dagger_{\downarrow \chi_2'}d^\dagger_{\uparrow \chi_1'}] F_{\uparrow \chi_1, \downarrow \chi_2} \overline{F}_{\uparrow \chi_1', \downarrow \chi_2'}
\end{equation}
with $\chi_1\chi_2=\chi_1'\chi_2'=\chi_0$ has a finite trace, there is no normal phase contraction if $\chi_1\ne\chi_1'$ and $\chi_2\ne\chi_2'$. As another example, in addition to superconductivity on the square lattice treated in Sec.~\ref{SubSec:UpdatesForSC}, consider superconductivity on an anisotropic triangular lattice with a 2x2 cluster in CDMFT. This cluster has $C_{2v}$ symmetry, and entries in the hybridization function $F$ with $\chi_0=A_2$ may be finite. Within the one particle basis, this happens with the irreducible representations $\chi_1=\chi_2'=B_1$ and $\chi_2=\chi_1'=B_2$ or $\chi_1=\chi_2'=B_2$ and $\chi_2=\chi_1'=B_1$. 
%In other words, the two operator updates are not ergodic as soon as there is more than one finite entry in $F_{\uparrow,\downarrow}$ transforming as the broken symmetry. 
%As in Sec. (\ref{SubSec:UpdatesForSC}), broken charge conservation allows to pair out the $F_{\uparrow,\downarrow}$ and $F_{\uparrow,\downarrow}^*$ entries in a contraction. The group of four operators defined by such a pair transforms as the identity, and the four operator updates restore ergodicity.\newline

The situation changes if only the spatial symmetry is broken, and entries in the hybridization $\Delta_{\sigma \chi_1, \sigma \chi_2}$ transforming as $\chi_0$ (i.e. $\overline{\chi}_1\chi_2=\chi_0$) are finite. Choose an $M>1$ such that $\chi_0^M = 1$. Then
\begin{equation}
\label{equ:SCConfig4}
\text{Tr}[d^\dagger_{\sigma \chi_1}d_{\sigma \chi_2}\cdots d^\dagger_{\sigma \chi_1}d_{\sigma \chi_2}]\Delta_{\sigma \chi_1, \sigma \chi_2}\cdots \Delta_{\sigma \chi_1, \sigma \chi_2}
\end{equation}
where $\Delta_{\sigma\chi_1, \sigma \chi_2}$ occurs $M$ times has finite weight but no normal phase contraction, since $\chi_1\ne\chi_2$ by definition. This means that two operator updates can never reach this configuration. 
%This is in contrast to the special case of a  superconducting problem where $\chi_1=\chi'_{1}$. 
In addition, insertion of more than four operators are necessary for ergodicity if $m > 2$, where $m$ is defined by the smallest non-zero integer such that $\chi_0^m = 1$.%\newline

To restore ergodicity, we begin by insertion and removal of operators as in equation (\ref{equ:SCConfig4}) with $M=m$. We have to include also all insertions and removals that come from other hybridzation functions $\Delta$ that transform as $\chi_0$, e.g. with some spins flipped. If $m=2$ this is sufficient. Otherwise $\chi_0\ne\overline{\chi}_0$, and there are two types of configurations which have to be considered. First, the configurations as in (\ref{equ:SCConfig4}), but for $\overline{\chi}_0^m$ as well. Second, configurations of the type $\chi_0\overline{\chi_0}$, analogue to equation (\ref{equ:SCConfig2}).%\newline

An example of a broken spatial symmetry with $m=2$ is anti-ferromagnetism. In the $\mathbf{K}$ basis of Sec. (\ref{SubSec:UpdatesForSC}), $\chi_0$ is the character corresponding to $(\pi,\pi)$. A possibility to avoid four operator updates here is to take the $C_{2v}$ group with mirror symmetry along the diagonals, as this symmetry is not broken.%\newline

Generalization to other broken symmetries and combinations of broken symmetries is straightforward, but may be tedious. Notice however, that the two operator updates are always ergodic whenever the cluster Hamiltonian is such that the trace can be evaluated in the segment representation.~\cite{Werner:2006, Werner:2006General,millisRMP} In that case creation and annihilation operators always come in pairs which transform as the identity. Otherwise the trace vanishes.

\section{Conclusion}\label{Sec:Discussion} 
While the use of symmetries of the cluster is a powerful tool to accelerate the evaluation of the trace over cluster states in the CTQMC hybridization solver, we have shown that the non-vanishing hybridization functions that arise in the presence of several classes of broken-symmetries in the bath generally introduce configurations of creation-annihilation operators in the cluster trace that cannot be reached with the usual updates that add or remove a pair of creation-annihilation operators. This phenomenon occurs with broken symmetries that involve spatial components. Ergodicity can be recovered by introducing updates with simultaneous insertion-removal of a larger numbers of pairs of creation-annihilation operators. Hamiltonians that lead to traces that can be evaluated in the segment algorithm~\cite{Werner:2006, Werner:2006General} are however exempt from this difficulty.    

As an example, we applied four operator updates that are necessary for ergodicity to the case of d-wave superconductivity in $2\times 2$ plaquette dynamical mean-field theory for the one-band Hubbard model. The results are qualitatively similar to those previously published,~\cite{Sordi:2012SC, Sordi:2013c} leading in particular to the same physical conclusions on the interplay between pseudogap and d-wave superconductivity. The results are however quantitatively better than previous ones. In particular, the range of doping over which superconductivity occurs close to $T=0$ is in better agreement with that found using the exact-diagonalization impurity solver.~\cite{kancharla} We thus expect that qualitative conclusions of previously published results using this algorithm for d-wave superconductivity~\cite{hauleDOPING,hauleCRITICAL,SentefSuperconductivity:2011,Sordi:2012SC, Sordi:2013c} will remain true, but the calculations should be revised for quantitative purposes. More importantly, one should keep in mind that in any new calculation in the presence of broken symmetries involving spatial components, one should include many-point updates in addition to the pair of creation-annihilation operator updates usually implemented. 

\begin{acknowledgments}
We are grateful to D. S\'en\'echal, G. Kotliar and K. Haule for useful discussions. This work has been supported by the Natural Sciences and Engineering Research Council of Canada (NSERC), and by the Tier I Canada Research Chair Program (A.-M.S.T.). Simulations were performed on computers provided by CFI, MELS, Calcul Qu\'ebec and Compute Canada.
\end{acknowledgments}

%\bibliography{sc}

%merlin.mbs apsrev4-1.bst 2010-07-25 4.21a (PWD, AO, DPC) hacked
%Control: key (0)
%Control: author (8) initials jnrlst
%Control: editor formatted (1) identically to author
%Control: production of article title (-1) disabled
%Control: page (0) single
%Control: year (1) truncated
%Control: production of eprint (0) enabled
%

\end{document}